\documentclass[aps,onecolumn]{revtex4}
\usepackage{graphicx}
\usepackage{subfigure}
\usepackage{caption2}
\usepackage{float}
\usepackage{amsmath}
\usepackage{amssymb}
\usepackage{color}

\begin{document}

\title{\textit{Alternate solitons}: Nonlinearly-managed one- and
two-dimensional solitons in optical lattices}
\author{Arthur Gubeskys, Boris A. Malomed, and Ilya M. Merhasin}
\affiliation{Department of Interdisciplinary Studies,\\
School of Electrical Engineering,\\
Faculty of Engineering, Tel Aviv University,\\
Tel Aviv 69978, Israel}

\begin{abstract}
We consider a model of Bose-Einstein condensates which combines a
stationary optical lattice (OL) and periodic change of the sign
of the scattering length (SL) due to the Feshbach-resonance
management. Ordinary solitons and ones of the gap type being
possible, respectively, in the model with constant negative and
positive SL, an issue of interest is to find solitons alternating,
in the case of the low-frequency modulation, between shapes of
both types, across the zero-SL point. We find such
\textit{alternate solitons} and identify their stability regions
in the 2D and 1D models. Three types of the dynamical regimes are
distinguished: stable, unstable, and semi-stable. In the latter
case, the soliton sheds off a conspicuous part of its initial norm
before relaxing to a stable regime. In the 2D case, the threshold
(minimum number of atoms) necessary for the existence of the
alternate solitons is essentially higher than its counterparts for
the ordinary and gap solitons in the static model. In the 1D
model, the alternate solitons are also found only above a certain
threshold, while the static 1D models have no threshold. In the 1D
case, stable antisymmetric alternate solitons are found too.
Additionally, we consider a possibility to apply the variational
approximation (VA) to the description of stationary gap solitons,
in the case of constant positive SL. It predicts the solitons in
the first finite bandgap very accurately, and does it reasonably
well in the second gap too. In higher bands, the VA predicts a
border between tightly and loosely bound solitons.
\end{abstract}

\maketitle

\section{Introduction}

Photonic lattices, i.e., effective periodic potentials created by
the interference of counter-propagating laser beams in
quasi-linear modes, provide for a powerful tool for the study of
nonlinear dynamics in other, nonlinear, modes in the same media. A
well-known example is a photonic lattice in a photorefractive
medium, where the beams launched in the ordinary polarization are
nearly linear, while the bias electric field applied to the
crystal makes the orthogonal extraordinary polarization strongly
nonlinear. This technique has made it possible to predict
\cite{prediction} and create \cite{Photorefr} various one- and
two-dimensional (1D and 2D) lattice solitons, including localized
vortices \cite{PhotorefrVortex}, in the extraordinary
polarization, supported by a harmonic photonic lattice in the
ordinary one.

Another physically significant example is provided by solitons in
Bose-Einstein condensates (BECs), that were predicted in the
presence of optical lattices (OLs) created by laser beams
illuminating the condensate (while the mean-field dynamics of the
BEC wave function is nonlinear due to effects of collisions
between atoms, the medium is completely linear for the optical
beams). In particular, 2D \cite{BBB1}-\cite{BBB3} and 3D
\cite{BBB1,BBB2} OLs were shown to stabilize solitons in BECs with
attractive interactions between atoms \cite{BBB1}-\cite{BBB3}
(without the lattice, the corresponding soliton solutions exist
too, but they are unstable against the spatiotemporal collapse
\cite{Berge}). Moreover, it has been demonstrated that
low-dimensional OLs, i.e., 1D and 2D lattice in the 2D
\cite{BBB2,BBB3} and 3D \cite{BBB2,BBB3,Barcelona}\ case,
respectively, can also stabilize fully localized multi-dimensional
solitons, giving them the freedom to move in the unrestricted
directions and thus collide \cite{BBB3}. A related result is the
demonstration of the stability of 2D \cite{Bessel} and 3D
\cite{Dumitru} solitons in the models with a cylindrical OL, that
can be induced by a diffraction-free Bessel beam (see Ref.
\cite{Bessel} and references therein).

If inter-atomic interactions in the BEC are repulsive (which is the most
common case \cite{Pethik}), solitons cannot exist in the free space, but
they can be created and supported, in the form of \textit{gap solitons}, by
the periodic OL potential \cite{Konotop,Ostrovskaya}. The 1D gap solitons
have been recently observed in the $^{87}$Rb condensate \cite{Oberthaler}.
Besides the fundamental solitons of the gap type, stable vortical solitons
have also been predicted in the 2D self-repulsive BECs trapped in the square
OL \cite{BBB2,Sakaguchi,Ostrovskaya2}.

Another mechanism which is was theoretically shown to be an
effective tool for the stabilization of 2D solitons in BECs (but
not of their 3D counterparts) is periodic temporal modulation of
the nonlinearity coefficient, which is possible through the
\textit{ac Feshbach resonance}. In this case, external magnetic
field alters the scattering length (SL)\ of the atomic collisions,
and the application of the ac field can periodically change the
sign of the interaction. This mechanism is the basis of the
so-called Feshbach-resonance-management (FRM) technique \cite{FRM}
(which may be regarded as an implementation of a general concept
of the \textit{nonlinearity management}). A possibility to employ
the FRM for the stabilization of 2D solitons -- in principle,
without any trapping field -- has been recently demonstrated
\cite{Fatkhulla,Japan,VPG} (in fact, this mechanism follows the
pattern of the one proposed earlier for the stabilization of
cylindrical optical solitons in a layered medium with the
periodically alternating sign of the Kerr coefficient
\cite{Isaac}). It is also relevant to mention that, while the 1D
OL, unlike its 2D and 3D counterparts, cannot stabilize 3D
solitons \cite{BBB3}, and the FRM technique per se does not make
it possible either, a combination of the FRM and
\emph{one-dimensional} OL readily stabilizes single- and
multi-peaked solitons in the 3D space \cite{Michal}.

Thus, two distinct kinds of solitons are possible in BECs, under stationary
conditions: ordinary ones in the case of self-attraction, and gap solitons
in the case of repulsion. The OL is necessary for the existence of the gap
solitons in any dimension, and for the stability of the ordinary ones,
except for the 1D case, where ordinary solitons are stable without the
lattice (only in this case ordinary solitons have been already observed in
the experiment \cite{experiment}). Solitons of the two kinds differ not only
by their existence and stability conditions, but also in the shape: ordinary
solitons normally have a single-peak structure without zeros, while the gap
solitons feature oscillatory tails with many zero-crossings.

The possibility of the application of the above-mentioned FRM
technique suggests to consider a ``mixed" situation, when the sign
of the nonlinearity periodically changes between the attraction
and repulsion. In the case of the high-frequency ``management",
the application of the averaging method makes this situation
tantamount to that in an effective static model \cite{Fatkhulla}.
However, in the quasi-adiabatic (low-frequency) case, one may
expect time-periodic alternation between solitons of the two
types. Such \textit{alternate solitons}, which were not considered
before, is the subject of the present work. In particular, an
issue is whether they are able to survive periodic passage of the
zero-SL point, as no soliton can exist in the static model without
nonlinearity. In this work, we chiefly focus on the 2D case, which
is the most interesting one; however, the 1D model is considered
too -- in particular, with the objective to compare conclusions
concerning the stability of alternate solitons in the different
dimensions, and the necessary conditions (\textit{threshold}) for
their existence.

Besides that, we also consider a more specific issue that was not studied
before, viz., a possibility to apply the well-known technique of the
variational approximation (VA) \cite{Progress} to stationary gap solitons in
OLs. Nevertheless, the main results reported in this work are based on
direct numerical simulations, as the VA technique for the alternate solitons
in the time-modulated system is too cumbersome.

The paper is organized as follows. In Section 2, we formulate the model,
including the consideration of its linear spectrum, and present results
generated by the VA for static 2D gap solitons on the square lattice. The
most essential results for the 2D alternate solitons are reported in Section
3, and respective results for the 1D model are given in Section 4. The paper
is concluded by Section 5. Some technical details of the VA are presented in
the Appendix.

\section{The model, linear spectrum, and variational approximation}

In the mean-field approximation, the evolution of the BEC\ single-atom wave
function $\psi $ in the presence of the 2D square-shaped OL and a tight trap
in the transverse (third) direction obeys an effective two-dimensional
Gross-Pitaevskii equation. Its normalized form is well known \cite{Pethik}:
\begin{equation}
i\frac{\partial \psi }{\partial t}+\frac{\partial ^{2}\psi
}{\partial x^{2}}+\frac{\partial ^{2}\psi }{\partial
y^{2}}+\varepsilon \left[ \cos (2x)+\cos (2y)\right] \psi +\lambda
(t)|\psi |^{2}\psi =0,  \label{model_2D}
\end{equation}where $\varepsilon $ is the strength of the square lattice, its spatial
period scaled to be $\pi $. In this work, we focus on solitons
which may exist in a large-area domain without the support from an
external parabolic-potential trap, therefore Eq. (\ref{model_2D})
does not include the latter term. The nonlinear coefficient
$\lambda (t)$ in Eq. (\ref{model_2D}) is proportional to the SL of
atomic collisions, which is controlled by the ac magnetic field
through the Feshbach resonance. The positive and negative signs of
$\lambda $ correspond to the self-attractive and repulsive
condensate, respectively. The only dynamical invariant of Eq.
(\ref{model_2D}) (with the time-dependent $\lambda (t)$) is the
norm, which is proportional to the number of atoms in the
condensate,\begin{equation} N=\int \int \left\vert \psi
(x,y,t)\right\vert ^{2}dxdy.  \label{N}
\end{equation}

In case of $\lambda =\mathrm{const}$, stationary solutions are
sought as $\psi (x,y,t)=u(x,y)\exp (-i\mu t)$ with a real chemical
potential $\mu $ and a real function $u$ satisfying the equation
\begin{equation}
\mu u+\frac{\partial ^{2}u}{\partial x^{2}}+\frac{\partial ^{2}u}{\partial
y^{2}}+\varepsilon \left[ \cos (2x)+\cos (2y)\right] u+\lambda u^{3}=0
\label{model_2D_stat}
\end{equation}The search for soliton solutions should be preceded by consideration of the
spectrum of the linearized version of Eq. (\ref{model_2D_stat}),
as solitons may only exist at values of $\mu $ belonging to gaps
in the spectrum. The linearization of Eq. (\ref{model_2D_stat})
leads to a separable 2D eigenvalue problem,\begin{equation} \left(
\hat{L}_{x}+\hat{L}_{y}\right) u(x,y)=-\mu u(x,y),
\label{operator}
\end{equation}where we have defined the 1D linear operator $\hat{L}_{x}\equiv \partial
^{2}/\partial x^{2}+\varepsilon \cos (2x)$. The corresponding
eigenstates can be built as $u_{kl}(x,y)=g_{k}(x)g_{l}(y)$, with
the eigenvalues $\mu _{kl}=\mu _{k}+\mu _{l}$, where $g_{k}(x)$
and $g_{l}(y)$ is any pair of quasi-periodic Bloch functions
solving the Mathieu equation, $\hat{L}_{x}g_{k}(x)=-\mu
_{k}g_{k}(x)$, $\mu _{k}$ and $\mu _{l}$ being the corresponding
eigenvalues. The band structure of the 2D linear equation
(\ref{operator}) constructed this way was already investigated in
detail (see, e.g., Refs. \cite{Ostrovskaya} and
\cite{DemetriMoti}). It includes, as usual, a semi-infinite gap
which extends to $\mu \rightarrow -\infty $, and a set of finite
gaps separated by bands that are populated by quasiperiodic
Bloch-wave solutions, see Figs. \ref{fig2} and \ref{fig3} below.

With the self-attractive nonlinearity ($\lambda >0$), a family of
stable stationary 2D solitons was found in the semi-infinite gap
\cite{BBB1,Yang,DemetriMoti}. With the repulsive nonlinearity,
$\lambda <0$, stable 2D gap-soliton solutions can be found in
finite gaps \cite{Konotop,Ostrovskaya,BBB2,Sakaguchi}. In either
case, a necessary condition for the existence of the stationary 2D
solitons is that their norm (see Eq. (\ref{N})) exceeds a certain
minimum (\textit{threshold}) value, $N_{\mathrm{thr}}$
\cite{DemetriMoti} (see some details below). Further, both the
ordinary and gap-mode solitons generated by Eq.
(\ref{model_2D_stat}) with $\lambda >0$ and $\lambda <0$,
respectively, can be categorized (following Ref. \cite{Sakaguchi})
as tightly-bound (TB) and loosely-bound (LB) ones. TB solitons are
essentially confined to a single cell of the OL potential, with
weak tails extending into adjacent cells (for this reason, similar
solutions were called \textit{single-cell solitons} in Ref.
\cite{BBB1}). On the contrary, LB solitons (alias
\textit{multi-cell} ones \cite{BBB1}) extend over many lattice
cells; examples of both types of the soliton's shape are given
below in Fig. \ref{fig4}.

The TB solitons are well approximated by a Gaussian \textit{ansatz}, which
is a basis for the application of the variational approximation (VA) to
them, as it was done in Refs. \cite{BBB1,BBB2,BBB3} for the self-attraction
case (the VA was first applied to BEC models in Refs. \cite{VA}). The
Lagrangian corresponding to the stationary equation (\ref{model_2D_stat}) is
\begin{equation}
L=\int_{-\infty }^{+\infty }\left\{ \mu u^{2}-\left[ \left(
\frac{\partial u}{\partial x}\right) ^{2}+\left( \frac{\partial
u}{\partial y}\right) ^{2}\right] +\frac{1}{2}\lambda
u^{4}+\varepsilon \left[ \cos (2x)+\cos (2y)\right] u^{2}\right\}
dx.  \label{L}
\end{equation}The Gaussian ansatz is taken in the form of\begin{equation}
u(x,y)=A\exp \left( -\frac{x^{2}+y^{2}}{2a^{2}}\right) ,  \label{Gauss}
\end{equation}which assumes that the soliton's center is pinned at a local potential
minimum, $x=y=0$ (it is a minimum provided that we set $\varepsilon >0$).
Substitution of the ansatz in the expression (\ref{L}) yields the effective
Lagrangian,
\begin{equation}
L_{\mathrm{eff}}=\pi A^{2}\left[ \mu a^{2}-1+\frac{\lambda
}{4}A^{2}a^{2}+2\varepsilon a^{2}\exp \left( -a^{2}\right) \right]
. \label{Leff}
\end{equation}This effective Lagrangian was already found in Ref. \cite{BBB1} and used, as
said above, to predict TB solitons in the case of the attractive
nonlinearity, with $\lambda =+1$ . Here, we do not fix $\lambda $; instead,
we fix the solution's norm (\ref{N}),
\begin{equation}
N_{\mathrm{sol}}=\pi A^{2}a^{2}\equiv 4\pi .  \label{fixedN}
\end{equation}

The variational equations following from the Lagrangian
(\ref{Leff}), $\partial L_{\mathrm{eff}}/\partial \left(
A^{2}\right) =\partial L_{\mathrm{eff}}/\partial \left(
a^{2}\right) =0$, along with the normalization condition
(\ref{fixedN}), lead to relations $A=2/a$ and
\begin{eqnarray}
\lambda  &=&1-2\varepsilon a^{4}\exp \left( -a^{2}\right) ,~  \label{a} \\
\mu  &=&-a^{-2}-2\varepsilon \left( 1-2a^{2}\right) \exp \left(
-a^{2}\right) ,  \label{mu}
\end{eqnarray}Equation (\ref{a}) with $\varepsilon >0$ generates two physical solutions
(ones with $a^{2}>0$) in the interval
\begin{equation}
1-8e^{-2}\varepsilon \leq \lambda \leq 1  \label{interval}
\end{equation}\cite{BBB1} (in the case of $\varepsilon =0$, the interval shrinks to the
single point $\lambda =1$, which precisely corresponds to the (unstable)
\textit{Townes soliton} of the free-space 2D nonlinear Schr\"{o}dinger
equation \cite{Berge}).

In the case of self-attraction, $\lambda >0$, the existence
interval (\ref{interval}) is present for any $\varepsilon $
\cite{BBB1}, and, for a relatively weak OL, with $\varepsilon
<\varepsilon _{\mathrm{cr}}=e^{2}/8\approx 0.92$, Eq.
(\ref{interval}) predicts a threshold condition for the existence
of the 2D soliton, $\lambda >\lambda _{\mathrm{thr}}\equiv
1-8e^{-2}\varepsilon $. The latter implies the existence of the
above-mentioned threshold value of the norm, if $N$ is allowed to
vary while $\lambda $ is fixed. For a strong lattice, with
$\varepsilon >\varepsilon _{\mathrm{cr}}$, Eq. (\ref{interval})
formally predicts zero threshold; however, as was shown in Ref.
\cite{BBB1}, the latter prediction is an artifact of the VA. In
reality, a finite $\lambda _{\mathrm{thr}}$ for fixed $N$, or,
equivalently, finite $N_{\mathrm{thr}}$ for fixed $\lambda $ is
always present. A \textit{delocalization transition} which happens
beneath the threshold was investigated in detail (including the
three-dimensional case) in Ref. \cite{Salerno}. Moreover, a
threshold necessary for the formation of solitons was also found
in the discrete version of the 2D model \cite{Kim}. Actually, the
threshold is found in the 2D model with any nonlinearity which can
support solitons, in combination with the lattice potential
\cite{DemetriMoti} (including 2D models with a quasi-1D potential
\cite{BBB2,BBB3}), while in the 1D models the threshold is absent.

For $\varepsilon >\varepsilon _{\mathrm{cr}}\equiv e^{2}/8$, Eq.
(\ref{interval}) predicts too that the solitons may exist in the
case of the repulsive nonlinearity, $\lambda <0$, as the
expression on the left-hand-side of Eq. (\ref{interval}),
$1-\varepsilon /\varepsilon _{\mathrm{cr}}$, is negative in this
case. Unlike the above-mentioned formally predicted zero threshold
for $\varepsilon >\varepsilon _{\mathrm{cr}}$ and $\lambda >0$,
this effect is a real one, as confirmed by numerical results
displayed below (in Figs. \ref{fig3} and \ref{fig4}). Figure
\ref{fig1} shows the chemical potential $\mu $ versus the
nonlinearity coefficient $\lambda $, as found from Eqs. (\ref{a})
and (\ref{mu}). As is seen, the solution curve indeed extends to
negative $\lambda \ $for $\varepsilon
>\varepsilon _{\mathrm{thr}}$, forming a loop.

The existence region for the 2D solitons, as predicted by the VA, is shown
in Fig. \ref{fig2}, against the backdrop of the band structure of the 2D
model. According to what was said above, two existence borders of the gap
solitons, both pertaining to the self-repulsion, $\lambda <0$, meet and
terminate, forming a cusp, at the above-mentioned value, $\varepsilon
=\varepsilon _{\mathrm{thr}}\approx 0.92$. The lower border very accurately
follows a narrow Bloch band separating the first finite gap from the
semi-infinite one, in which the ordinary solitons (corresponding to the
self-attraction, $\lambda >0$) are predicted to exist. On the other hand,
the upper existence border predicted by the VA for the gap solitons cannot
be accurate since, in higher finite gaps, the actual soliton's shape (see
examples in Fig. \ref{fig4} below) is very far from the Gaussian ansatz
adopted above. However, comparison with numerically found shapes of the
solitons demonstrates that this upper border gives, as a matter of fact,
quite an accurate approximation for a separatrix between regions of tightly
and loosely bound (TB and LB) solitons in the repulsion case.

Fixing the OL's strength at $\varepsilon =7.5$, we solved the stationary
equation (\ref{model_2D_stat}) numerically and compared the results with the
VA prediction, as shown in Figs. \ref{fig3} and \ref{fig4}. We observe that,
in the semi-infinite and first finite gaps, the accuracy of the prediction
for both $\lambda >0$ and $\lambda <0$ (attraction and repulsion) is very
good, and in the second finite gap it is fair. The third finite gap is
located just above the VA-predicted upper border of the existence region for
the gap solitons (the upper dashed line in Fig. \ref{fig2}). The solitons
found numerically in the third gap, see panel (f) in Fig. \ref{fig4}, are
definitely of the LB type, and they cannot be approximated by the Gaussian
ansatz.

We tried to apply the VA to loosely bound (LB) solitons, introducing an
ansatz that allows oscillatory tails, cf. Fig. \ref{fig4}(f):
\begin{equation}
u(x,y)=\left[ A+B\left( \cos x+\cos y\right) \right] \exp \left(
-\frac{x^{2}+y^{2}}{2a^{2}}\right) .  \label{cos}
\end{equation}Cumbersome variational equations generated by the substitution of this
ansatz in the Lagrangian (\ref{L}) are given in the Appendix. In
Fig. \ref{fig5} we compare the LB-solitons' shapes, as predicted
by the extended ansatz and found from the direct numerical
solution of Eq. (\ref{model_2D_stat}). As is seen in panel (a),
the new ansatz fails to adequately describe sidelobes of a soliton
which is intermediate between the TB and LB types. For the LB
soliton proper, panel (b) shows that three central peaks are
correctly approximated by the modified VA, but the discrepancy is
large farther from the center.

\section{Dynamics of two-dimensional solitons under the Feshbach-resonance
management}

In this section we address the main issue of the work, viz.,
solitons driven by the FRM. The consideration is based on direct
simulations of Eq. (\ref{model_2D}) with the norm fixed as per Eq.
(\ref{fixedN}) and the time-dependent nonlinear coefficient,
\begin{equation}
\lambda (t)=\lambda _{0}+\lambda _{1}\cos (\omega t).  \label{lambda(t)}
\end{equation}We begin with the case of the vanishing dc part, $\lambda _{0}=0$. At $t=0$,
we use the initial profile corresponding to a numerically found soliton
solution of the stationary equation (\ref{model_2D_stat}) with $\lambda
=\lambda _{1}$ (we set $\lambda _{1}>0$). Systematic simulations demonstrate
that it is possible to achieve \emph{stable} periodic adiabatic alternations
between two quasi-stationary soliton shapes, one corresponding to an
ordinary soliton belonging to the semi-infinite gap in the case of the
attractive nonlinearity, and the other being a gap soliton in one of the
finite gaps, which exists with the repulsive nonlinearity. Relaxation of
this \textit{alternate soliton} to a stable regime is accompanied by very
weak radiation loss. In the simulations of the 2D case, we incorporated
absorbers near boundaries of the integration domain, in order to emulate an
infinite space.

An example of a robust alternate soliton is given in Fig.
\ref{fig6}. In particular, sidelobes in the soliton's profile,
characteristic of the gap-soliton shape, periodically appear and
disappear. It is noteworthy that periodic crossings of the zero-SL
point, $\lambda =0$, at which no stationary soliton may exist, do
not destroy the alternate soliton. The spatially-averaged squared
width of the soliton, the temporal evolution of which is shown in
the lower panel of the figure, is defined as\begin{equation} \xi
^{2}(t)\equiv \frac{\int \int x^{2}|u(x,y,t)|^{2}dxdy}{\int \int
|u(x,y)|^{2}dxdy}.  \label{xi}
\end{equation}

Results of systematic simulations are summarized in stability
diagrams for the alternate solitons, which are displayed in Fig.
\ref{fig7} for $\lambda_0=0$ and several different values of the
FRM amplitude $\lambda_1$. Naturally, the solitons may be stable
in the case of the quasi-adiabatic FRM driving, i.e., at
sufficiently low frequencies. In the stability region, the total
radiation loss is less than $2\%$ of the initial norm (number of
atoms), which is our definition of complete robustness of the
alternate solitons. In particular, the example shown in Fig.
\ref{fig6} corresponds to the point (a) in Fig. \ref{fig7} (for
$\lambda_1=0.7$); in this case, the total loss is almost exactly
$2\%$.

As the driving frequency $\omega $ increases, the soliton emits
more radiation. For moderately high frequencies, the initial
solitary-wave pulse prepared as said above (i.e., as a numerically
exact stationary soliton corresponding to the initial value of
$\lambda $) sheds off a conspicuous share of its norm; then, the
emission of radiation ceases, and the remaining part of the pulse
self-traps into a robust alternate soliton. An example of a such a
\textit{semi-stable} dynamical regime is displayed in Fig.
\ref{fig8}. To additionally illustrate the relaxation to the
stable regime, the lower panel of the figure includes a plot
showing the evolution of the norm,\begin{equation} \rho
(t)=\int \int |u(x,y,t)|^{2}dxdy, \label{rho}
\end{equation}cf. Eq. (\ref{xi}). In this case, the resulting alternate soliton oscillates
between near-stationary shapes corresponding to points (a) and (b)
in Fig. \ref{fig3}, which belong to the semi-infinite and first
finite gaps, respectively. In general, the stronger the OL is, the
more robust the alternate soliton will be. In Fig. \ref{fig7}, the
semi-stable regimes are not marked separately from completely
unstable ones, as the border between them is fuzzy (in particular,
it is not quite clear whether the semi-stable solitons would not
very slowly decay on an extremely long time scale, unavailable to
simulations that we could run). In any case, a broad area adjacent
to the one marked as completely stable one in Fig. \ref{fig7} is
actually a region of semi-stability. At still higher driving
frequencies, the soliton is definitely destroyed, see a typical
example in Fig. \ref{fig9}.

It is not quite clear either if the stability region of the
alternate solitons is limited on the side of very small
frequencies. Indeed, one may expect that, in the latter case, the
soliton spending long time around the zero-SL point, $\lambda =0$,
must spread out, and may thus decay; on the other hand, if the
soliton is very broad by itself, it may survive this temporary
spreading out. The lowest frequency checked in our simulations was
$\omega =0.1$, the solitons being unequivocally stable at this
value of $\omega $. Still lower frequencies require impractically
long simulation times (and extremely large simulation domains).
Accumulation of numerical error could impede to draw certain
conclusions in this case. Extremely long evolution and very large
domains are not relevant either from the viewpoint of experiments
with BECs.

It is also relevant to address the issue of the existence of the
threshold necessary for the formation of the soliton, which, as
explained above, exists for both $\lambda >0$ and $\lambda <0$ in
the static 2D models. In the nonstationary (FRM-driven) model with
the fixed norm (see Eq. (\ref{fixedN})), the threshold manifests
itself in the fact that persistent alternate solitons cannot be
found if the nonlinearity coefficient is too small, $\lambda
_{1}<\left( \lambda _{1}\right) _{\mathrm{thr}}$. Identifying the
threshold directly from simulations of Eq. (\ref{model_2D}) is
difficult (as well as locating other precise borders of the
existence and stability regions for the alternate solitons, see
above). For instance, for the strong lattice with $\varepsilon
=7.5$ (cf. the situation for the same value of $\varepsilon $ in
the stationary model, as illustrated by Fig. \ref{fig3}), the
threshold exists but is so small that its accurate value cannot be
identified. This is possible for smaller $\varepsilon $. In
particular, for $\varepsilon =4$ we have found $\left( \lambda
_{1}\right) _{\mathrm{thr}}\approx 0.15$, which should be compared
to the the thresholds found, with the same $\varepsilon =4$ and
the same fixed norm (\ref{fixedN}), for the ordinary and gap
solitons in the corresponding static 2D models: $\lambda
_{\mathrm{thr}}^{\mathrm{(ord)}}\approx 0.04$, and $\lambda
_{\mathrm{thr}}^{\mathrm{(gap)}}\approx -0.04$, respectively.
Quite naturally, the dynamical threshold is much higher than its
static counterparts; at $\lambda _{1}<\left( \lambda _{1}\right)
_{\mathrm{thr}}$, the alternate soliton clearly demonstrates a
delocalization transition.

Next, we consider the FRM-driven soliton dynamics with a
\emph{negative} nonzero dc part in Eq. (\ref{lambda(t)}), $\lambda
_{0}<0$, which corresponds to repulsion. To this end, we consider
an example with $\lambda _{0}=-0.9$ and $\lambda _{1}=1.6$. The
interaction being repulsive on the average, one may expect the
existence of gap solitons in the high-frequency limit. In the
simulations, we started with the initial profile corresponding to
point (a) in Fig. \ref{fig3}, as the initial value of the
nonlinearity coefficient, $\lambda (0)=0.7$, pertains to this
point. The minimum instantaneous value of the oscillating
nonlinear coefficient is $\lambda _{\min }=-2.5$ in the present
case, the stationary solution with $\lambda =-2.5$ pertaining to
point (d) in Fig. \ref{fig3}, which belongs to the second finite
band. In this regime, the oscillating nonlinear coefficient
$\lambda (t)$, in addition to cycling across the point $\lambda
=0$ and a very narrow Bloch band separating the semi-infinite and
first finite gaps, periodically passes the wider Bloch band
between the first and the second finite gaps, where stationary
solitons cannot exist. Nevertheless, a stable alternate soliton,
found in this case, survives all the traverses of the ``dangerous
zones", as shown in Fig. \ref{fig10}. A small amount of radiation
is emitted at an initial stage of the evolution, and then a robust
alternate soliton establishes itself, cf. Fig. \ref{fig8}. It is
noteworthy that, as seen in the insets, this soliton develops
sidelobes that actually do not oscillate together with its core,
and do not disappear either as $\lambda $ takes positive values.
The latter feature distinguishes this stable regime from the one
shown in Fig. \ref{fig6}, where the sidelobes periodically
disappear.

We have also tried to apply the FRM mechanism to weakly localized
LB solitons, such as the one in panel (f) of Fig. \ref{fig4}.
However, no stable regime periodically passing through a shape of
this type could be found for any combination of $\lambda _{0}$ and
$\lambda _{1}$ in Eq. (\ref{lambda(t)}). A typical example of
unstable evolution of FRM-driven LB solitons is displayed in Fig.
\ref{fig11}.

\section{Dynamics of one-dimensional solitons under the Feshbach-resonance
management}

The 1D case also deserves consideration, as the experiment may be
easier in this case, and it is interesting to compare the results
with those reported above for the 2D model. In particular, an
issue is whether the existence of the 1D alternate soliton entails
any threshold condition. It is well known that the effective 1D
Gross-Pitaevskii equation is a straightforward reduction of Eq.
(\ref{model_2D}),\begin{equation} i\frac{\partial \psi }{\partial
t}+\frac{\partial ^{2}\psi }{\partial x^{2}}+\varepsilon \cos
(2x)\psi +\lambda (t)|\psi |^{2}\psi =0,  \label{1D}
\end{equation}which implies that a tight trap acts in the directions $y$ and $z$. Here we
report results only for the (most fundamental) case without the dc component
in $\lambda (t)$, i.e., $\lambda _{0}=0$, and fixing $\left\vert \lambda
_{1}\right\vert =1$ in Eq. (\ref{lambda(t)}). The strategy is the same as in
the 2D case: direct simulations of Eq. (\ref{1D}) start with a soliton
profile that would be a numerically exact stationary soliton for the initial
value of the nonlinearity coefficient, $\lambda =\lambda (0)$. In most
cases, the solution's 1D norm was fixed at $N\equiv \int_{-\infty }^{+\infty
}\left\vert u(x)\right\vert ^{2}dx=7.9$ (this normalization was chosen as it
corresponds to an almost constant value of the chemical potential, $\mu
\approx 2$, in the stationary version of the problem). However, the overall
stability diagram will include different values of $N$, see Fig. \ref{fig15}
below.

In the 1D case, stable alternate solitons can be readily found,
see an example in Fig. \ref{fig12}. The soliton periodically
oscillates between the narrow and wide profiles, which are
displayed in Fig. \ref{fig13}. Taking a smaller OL's strength
$\varepsilon $ and/or larger frequency $\omega $, we observe onset
of strong instability of the soliton, as shown in Fig.
\ref{fig14}.

Collecting results of the systematic simulations, we have
generated a stability diagram for the 1D alternate solitons, which
is displayed in Fig. \ref{fig15}. It is noteworthy that the shape
of the stability area is qualitatively similar to that for the 2D
case, cf. Fig. \ref{fig7}. The similarity suggest that the basic
results for the stability of the alternate solitons are quite
generic.

As is well known, in the static 1D models with both $\lambda >0$
and $\lambda <0$, the existence of the ordinary and gap solitons
\emph{does not} require any finite threshold (minimum norm). A
principal difference of the dynamic (FRM-driven) 1D model is that
persistent alternate solitons can be found only \emph{above a
finite threshold}, $N>N_{\mathrm{thr}}$. Accurate identification
of $N_{\mathrm{thr}}$ is rather difficult but possible. For
instance, we have found $N_{\mathrm{thr}}\approx 2.1$ for
$\varepsilon =7.5$ and $\omega =\pi /2$. Thus, in this sense the
1D dynamic model is closer to the 2D one than to its static 1D
counterparts. It remains to investigate the existence of the
finite threshold in the 1D dynamic model with a nonzero dc
component, i.e., $\lambda _{0}\neq 0$ in Eq. (\ref{lambda(t)}).

Besides the fundamental (single-peaked) 1D solitons considered above, stable
higher-order (multi-peaked) alternate solitons have been found too. As
concerns static higher-order solitons on lattices, a known example is the
so-called intrinsic localized mode, i.e., an odd (antisymmetric) soliton in
the discrete nonlinear Schr\"{o}dinger equation \cite{ILM}. A similar object
is a bound state of two lattice solitons, which, too, is stable only in the
anti-symmetric configuration, in the 1D \cite{Todd} and 2D \cite{Bishop}
cases alike.

Following the pattern of the static lattice models, we prepared an
antisymmetric initial state as a superposition of two separated solitons
(stationary ones corresponding to the initial value of $\lambda $) with
opposite signs, i.e., the phase difference of $\pi $. Direct simulations
demonstrate that stable alternate antisymmetric solitons can be easily found
this way, see a typical example in Fig. \ref{fig16}. Stable solitons of
still higher orders, i.e., bound states of several fundamental solitons with
the phase shift $\pi $ between them, were found too.

A 2D counterpart of the odd soliton would be a vortical soliton.
Such stable vortices were found indeed in the static models with
self-attraction \cite{BBB1,Yang} and self-repulsion
\cite{BBB2,Sakaguchi,Ostrovskaya2}. However, our simulations have
not produced stable solitons with intrinsic vorticity in the 2D
FRM lattice model based on Eqs. (\ref{model_2D}) and
(\ref{lambda(t)}), with $\lambda _{0}=0$.

\section{Conclusion}

We have introduced a model of Bose-Einstein condensates in a
system combining the optical lattice (OL) with the strength
$\varepsilon ~$and periodic modulation of the scattering length
(SL) with the frequency $\omega $, which is provided by external
ac magnetic field through the Feshbach resonance. As it was known
before that stationary ordinary solitons and ones of the gap type
are possible in the model with constant negative and positive SL,
respectively, an issue of interest is to find a stable soliton
periodically alternating, in the case of the low-frequency
modulation, between shapes of both types. In both 2D and 1D cases,
we have found such \textit{alternate solitons}, and identified
their stability regions in the $\left( \varepsilon ,\omega \right)
$ plane. As might be expected, the stability regions are limited
to relatively low frequencies and sufficiently strong OL. A
threshold necessary for the formation of the alternate solitons
can be found in the 2D model, and also in its 1D counterpart (with
the zero mean value of the SL); the latter result is especially
interesting, as the 1D static models have no threshold. Dynamical
regimes of three types were identified: stable, unstable, and
semi-stable, the latter one implying that the initial soliton
sheds off a conspicuous part of its norm before relaxing into a
stable regime. In the 1D case, stable antisymmetric alternate
solitons were additionally found, while stable vortical solitons
were not generated by simulations of the 2D model with the zero
average SL.

We have also considered a possibility to apply the variational approximation
(VA) to the description of stationary gap solitons in the 2D model with a
constant positive SL (self-repulsion). We have concluded that the Gaussian
ansatz very accurately predicts the solitons in the first finite bandgap,
and predicts them with a fair accuracy in the second gap. In higher gaps,
the VA actually produces a border between tightly and loosely bound
solitons. We also tried to apply a more sophisticated version of the VA,
based on the ansatz combining the Gaussian and cosines, with the objective
to describe the shape of loosely bound 2D gap solitons with oscillatory
tails. It was found that the modified ansatz can correctly describe three
central peaks in the soliton's shape.

\section*{Acknowledgement}

This work was supported, in a part, by the Israel Science Foundation through
the grant No. 8006/03.

\section*{Appendix: The variational approximation with the extended ansatz}

The effective Lagrangian generated by the substitution of the modified
ansatz (\ref{cos}) in Eq. (\ref{L}) is
\begin{multline}
L_{\mathrm{eff}}=\frac{\pi a^{2}}{16}e^{-4a^{2}}\left\{
8B^{3}Ae^{23a^{2}/8}{\left( 1+3e^{a^{2}/2}\right) }^{2}\lambda
+B^{4}e^{2a^{2}}{\left(
1+e^{a^{2}/4}\right) }^{4}{\left( 1-2e^{a^{2}/4}+3e^{a^{2}/2}\right) }^{2}\lambda +\right.  \\
32BAe^{7a^{2}/4}\left[ {\left( 1+e^{a^{2}}\right) }^{2}\varepsilon
+e^{2a^{2}}\left( A^{2}e^{a^{2}/8}\lambda +2\mu \right) \right]
+4A^{2}e^{3a^{2}}\left[ 8\varepsilon +e^{a^{2}}\left( A^{2}\lambda +4\mu
\right) \right] + \\
\left. 8B^{2}\left( {\left( 1+2e^{3a^{2}/2}+e^{2a^{2}}\right)
}^{2}\varepsilon +e^{3a^{2}}\left[ 3A^{2}e^{a^{2}/2}{\left(
1+e^{a^{2}/4}\right) }^{2}\lambda +2{\left( 1+e^{a^{2}/2}\right)
}^{2}\mu \right] \right) \right\} .
\end{multline}

The variation with respect to the parameters $B$, $a$ and $A$ yields the
following equations:
\begin{subequations}
\begin{multline}
B^{3}e^{2a^{2}}{\left( 1+e^{a^{2}/4}\right) }^{4}{\left(
1-2e^{a^{2}/4}+3e^{a^{2}/2}\right) }^{2}\lambda + \\
6Ae^{23a^{2}/8}{\left( B+3Be^{a^{2}/2}\right) }^{2}\lambda
+8Ae^{7a^{2}/4}\left[ {\left( 1+e^{a^{2}}\right) }^{2}\varepsilon
+e^{2a^{2}}\left(
A^{2}\lambda e^{a^{2}/8}+2\mu \right) \right] + \\
4B\left\{ {\left( 1+2e^{3a^{2}/2}+e^{2a^{2}}\right) }^{2}\varepsilon
+e^{3a^{2}}\left[ 3A^{2}e^{a^{2}/2}{\left( 1+e^{a^{2}/4}\right) }^{2}\lambda
+2{\left( 1+e^{a^{2}/2}\right) }^{2}\mu \right] \right\} =0,
\end{multline}\begin{multline}
-8\left( -1+4a^{2}\right) B^{2}\varepsilon -16\left( -2+5a^{2}\right)
B^{2}e^{3a^{2}/2}\varepsilon -8\left( -4+9a^{2}\right)
BAe^{7a^{2}/4}\varepsilon - \\
\left( -8+9a^{2}\right) B^{3}Ae^{23a^{2}/8}\lambda -6\left( -8+5a^{2}\right)
B^{3}Ae^{27a^{2}/8}\lambda - \\
\left( -8+a^{2}\right) BA\left( 9B^{2}+4A^{2}\right) e^{31a^{2}/8}\lambda
-\left( -1+2a^{2}\right) B^{2}e^{2a^{2}}\left( 16\varepsilon +B^{2}\lambda
\right) - \\
2\left( -4+5a^{2}\right) Be^{11a^{2}/4}\left( 8A\varepsilon +B^{3}\lambda
\right) -4\left( -2+a^{2}\right) B^{2}e^{7a^{2}/2}\left( 4\varepsilon
+2B^{2}\lambda +3A^{2}\lambda +4\mu \right) - \\
2\left( -1+a^{2}\right) e^{3a^{2}}\left[ 16A^{2}\varepsilon +3B^{4}\lambda
+8B^{2}\left( 2\varepsilon +\mu \right) \right] - \\
2\left( -4+a^{2}\right) Be^{15a^{2}/4}\left[ 3B^{3}\lambda +6BA^{2}\lambda
+4A\left( \varepsilon +2\mu \right) \right] + \\
e^{4a^{2}}\left[ 9B^{4}\lambda +8B^{2}\left( \varepsilon +3A^{2}\lambda
+2\mu \right) +4A^{2}\left( A^{2}\lambda +4\mu \right) \right] =0
\end{multline}\begin{multline}
6B^{2}Ae^{7a^{2}/4}{\left( 1+e^{a^{2}/4}\right) }^{2}\lambda
+B^{3}e^{9a^{2}/8}{\left( 1+3e^{a^{2}/2}\right) }^{2}\lambda
+2Ae^{5a^{2}/4}\left[ 4\varepsilon +e^{a^{2}}\left( A^{2}\lambda +2\mu \right) \right]  \\
+4B\left[ {\left( 1+e^{a^{2}}\right) }^{2}\varepsilon +e^{2a^{2}}\left(
3A^{2}e^{a^{2}/8}\lambda +2\mu \right) \right] =0.
\end{multline}Examples of results produced by the solution of these equations are
displayed in Fig. \ref{fig5}.

\newpage

\begin{figure}[tbp]
\includegraphics[scale=1]{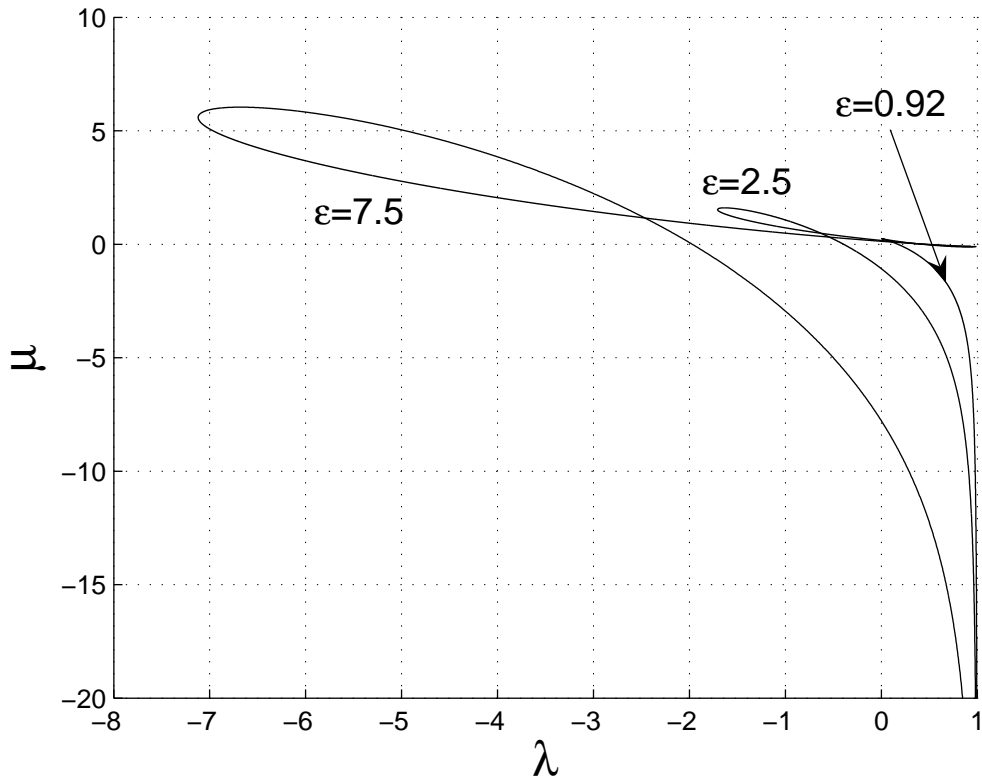}
\caption{The chemical potential $\protect\mu $ vs. the nonlinearity
coefficient $\protect\lambda $ for the soliton family, as given by the
variational approximation for different values of the optical-lattice
strength $\protect\varepsilon $.}
\label{fig1}
\end{figure}

\begin{figure}[tbp]
\includegraphics[scale=1]{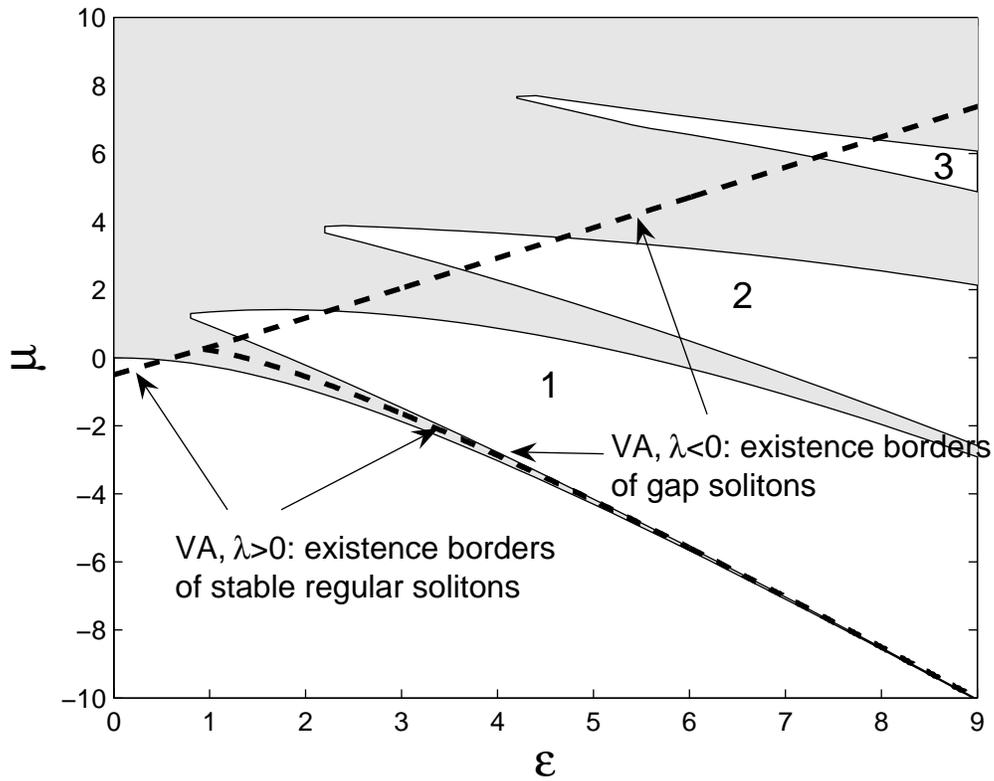}
\caption{The band-gap structure and regions of the existence of
regular ($\protect\lambda >0$) and gap ($\protect\lambda <0$)
solitons, as predicted by the variational approximation based on
the Gaussian ansatz. Shaded and unshaded zones are, respectively,
the Bloch bands (where solitons cannot exist) and gaps (where
solitons are possible). Dashed lines are borders of the soliton
existence regions.} \label{fig2}
\end{figure}

\begin{figure}[tbp]
\includegraphics[scale=1]{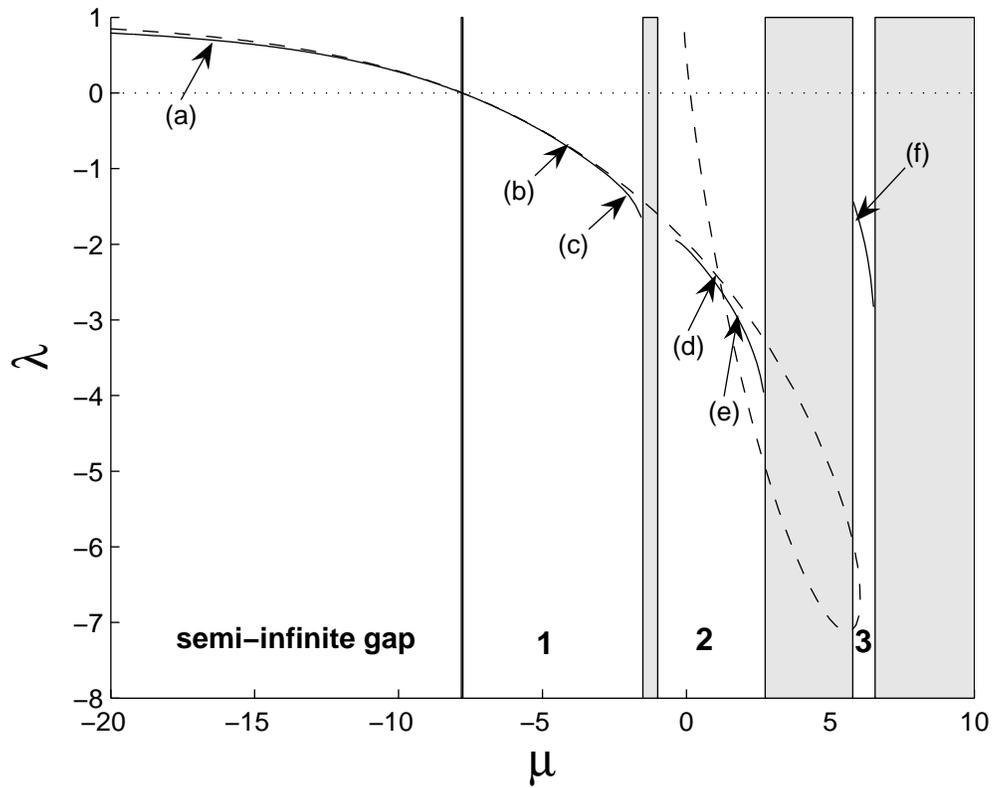}
\caption{Comparison of the numerically generated soliton family
and the one predicted by the variational approximation (solid and
dashed lines, respectively) for the fixed value of the
optical-lattice's strength, $\protect\varepsilon =7.5$. Shaded and
unshaded zones are the Bloch spectral bands and gaps,
respectively.} \label{fig3}
\end{figure}

\begin{figure}[tbp]
\includegraphics[scale=1]{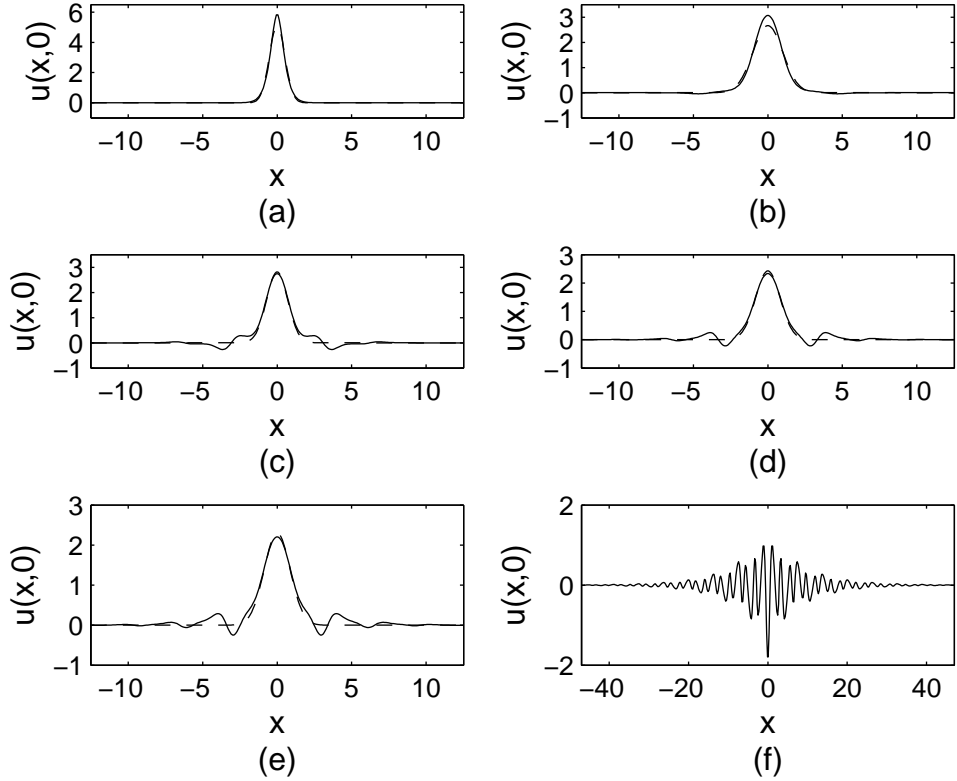}
\caption{Comparison of the soliton profiles found numerically and
those generated by the variational approximation (solid and dashed
lines, respectively) for the same fixed value of the
optical-lattice's strength as in Fig. \protect\ref{fig3},
$\protect\varepsilon =7.5$. Plots (a) to (f) correspond to points
in the parameter plane ($\protect\mu $, $\protect\lambda $)
labeled by the same symbols in Fig. \protect\ref{fig3}. This
figure and Fig. \protect\ref{fig5} show cross sections of the
two-dimensional solitons along the axis $y=0$.} \label{fig4}
\end{figure}

\begin{figure}[tbp]
\subfigure[]{\includegraphics[width=2.5in]{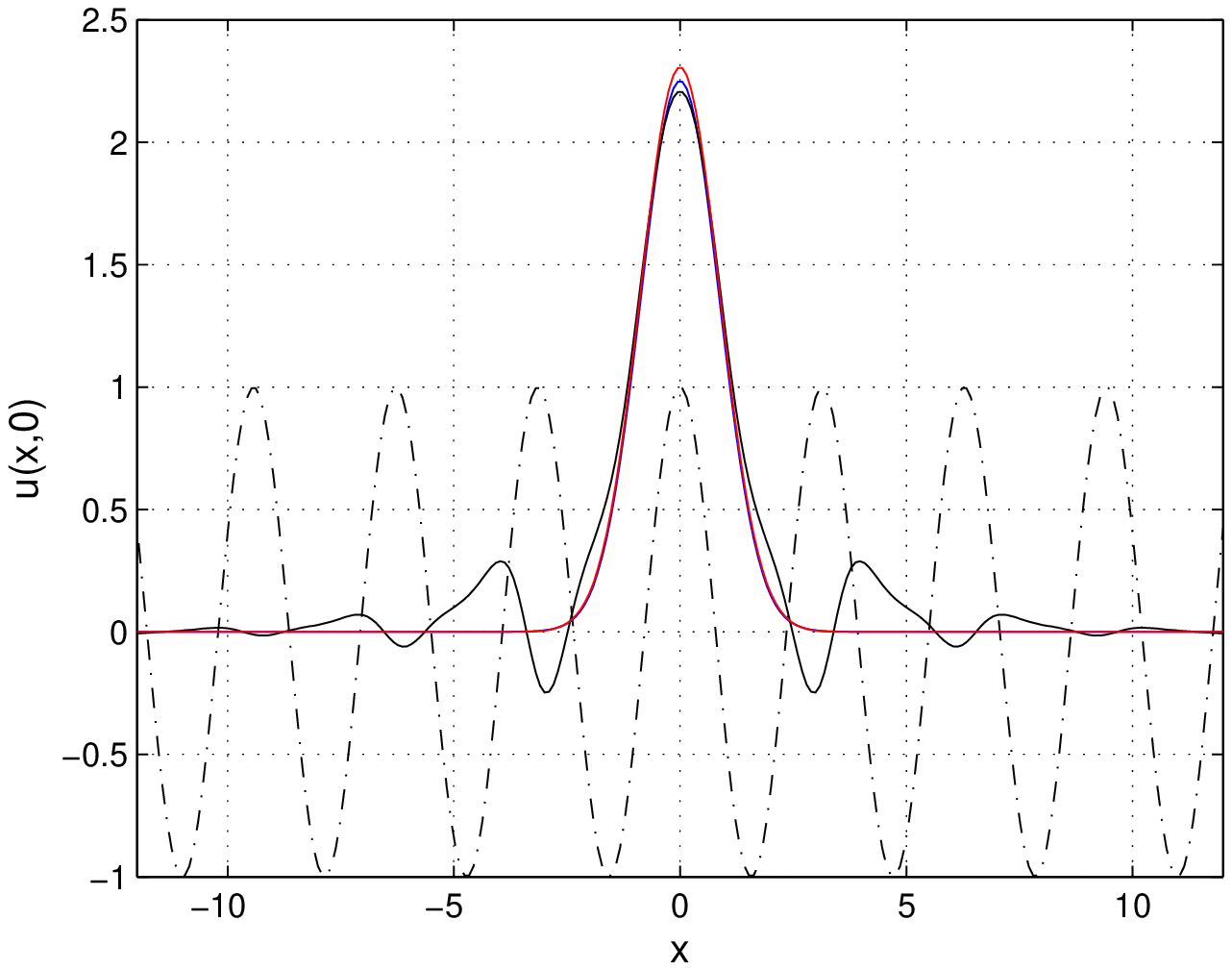}}
\subfigure[]{\includegraphics[width=2.5in]{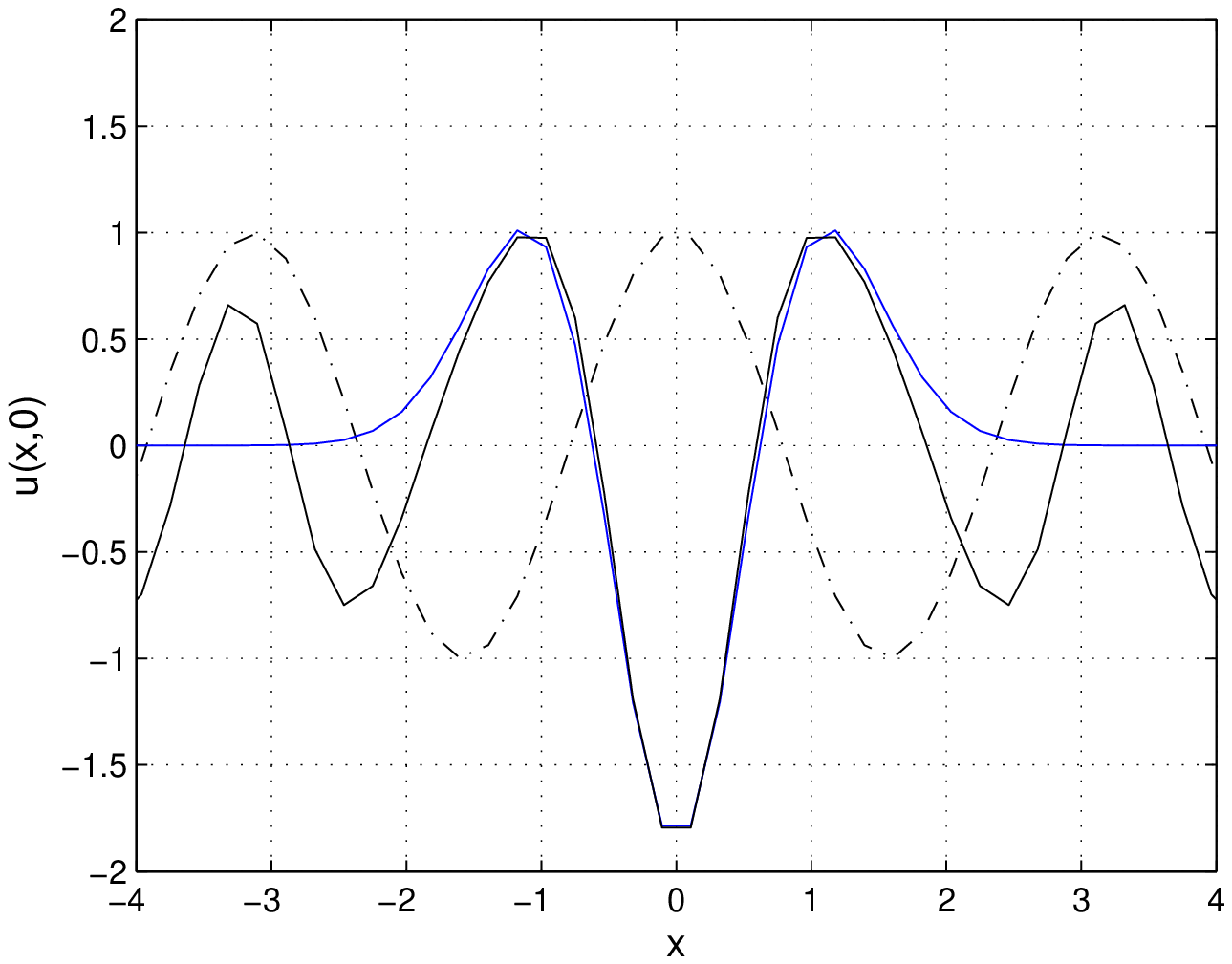}}
\caption{Comparison of profiles generated by the variational
approximation for the stationary two-dimensional solitons based on
the simple Gaussian ansatz (\protect\ref{Gauss}) (red line),
extended ansatz (\protect\ref{cos}) (blue line), and direct
numerical results (black line). The dashed-dotted curve shows the
periodic potential. (a) A case intermediate between tightly- and
loosely-bound solitons, corresponding to panel (e) in Fig.
\protect\ref{fig4}. (b) Zoom into the loosely-bound-soliton's
profile corresponding to panel (f) in Fig. \protect\ref{fig4} (in
this case, the profile generated by the simple ansatz
(\protect\ref{Gauss}) is not included, as it is completely
irrelevant).} \label{fig5}
\end{figure}

\begin{figure}[tbp]
\includegraphics[scale=1]{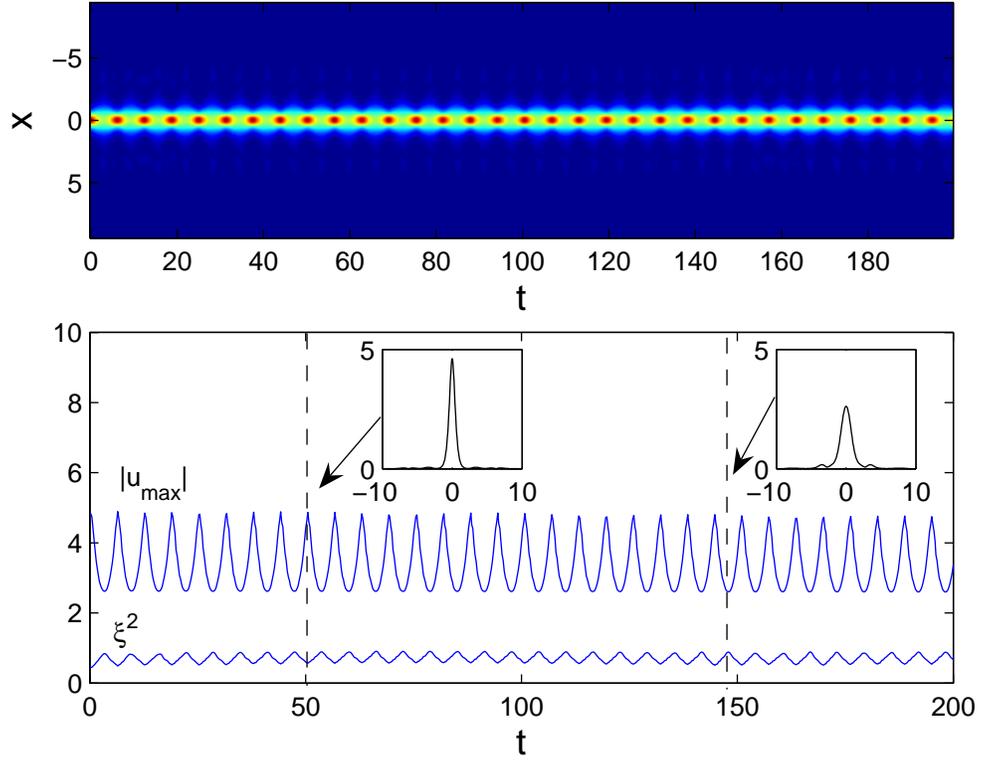}
\caption{An example of an almost completely stable alternate
soliton, for $\protect\varepsilon =5$, $\protect\lambda _{1}=0.7$,
$\protect\omega =1$, and $\protect\lambda _{0}=0$; it corresponds
to point (a) in Fig. \protect \ref{fig7} (for $\protect\lambda
_{1}=0.7$). The upper panel shows the soliton evolution in terms
of contour plots. The lower panel shows the time dependence of the
amplitude and mean squared width of the soliton. Two insets are
cross-sections of instantaneous profiles of the soliton taken at
moments of time ($t=50$ and $t=150$, respectively) when it is very
close, respectively, to a regular soliton and one of the gap
type.} \label{fig6}
\end{figure}

\begin{figure}[tbp]
\includegraphics[scale=1]{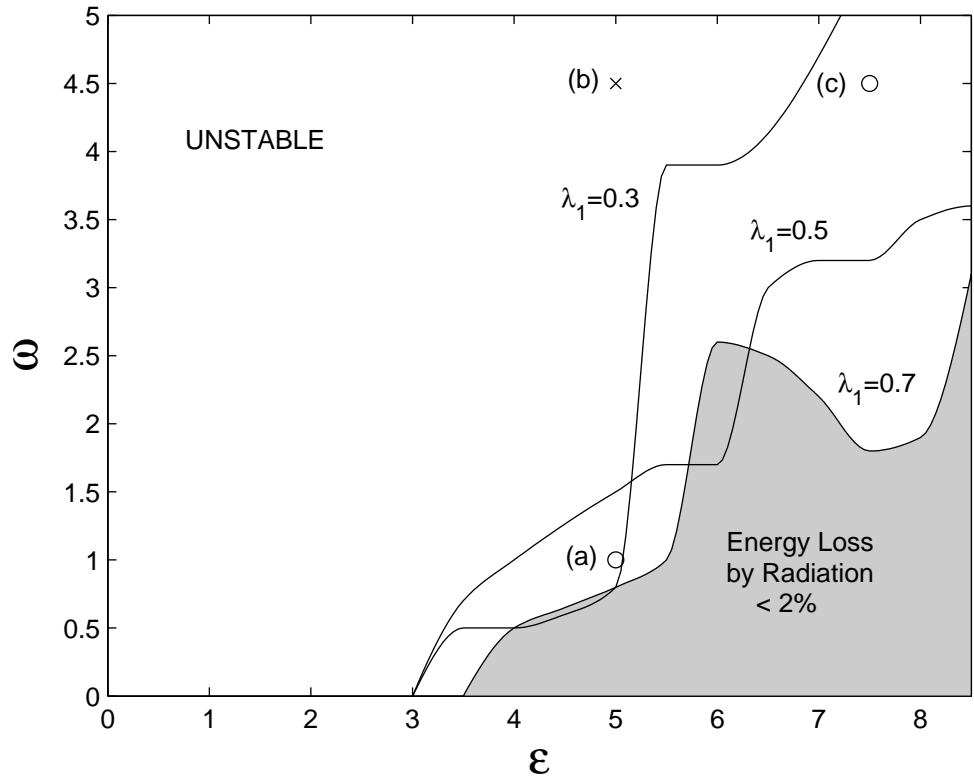}
\caption{The stability diagram of the alternate solitons in the
($\protect\varepsilon $,$\protect\omega $) plane for
$\protect\lambda _{0}=0$ and different fixed values of the
Feshbach-resonance-management amplitude $\protect\lambda _{1}$.
The region of complete stability (it is shaded for
$\protect\lambda _{1}=0.7$) is defined so that the total radiation
loss of the soliton's initial norm is less than $2\%$ in this
region.} \label{fig7}
\end{figure}

\begin{figure}[tbp]
\includegraphics[scale=1]{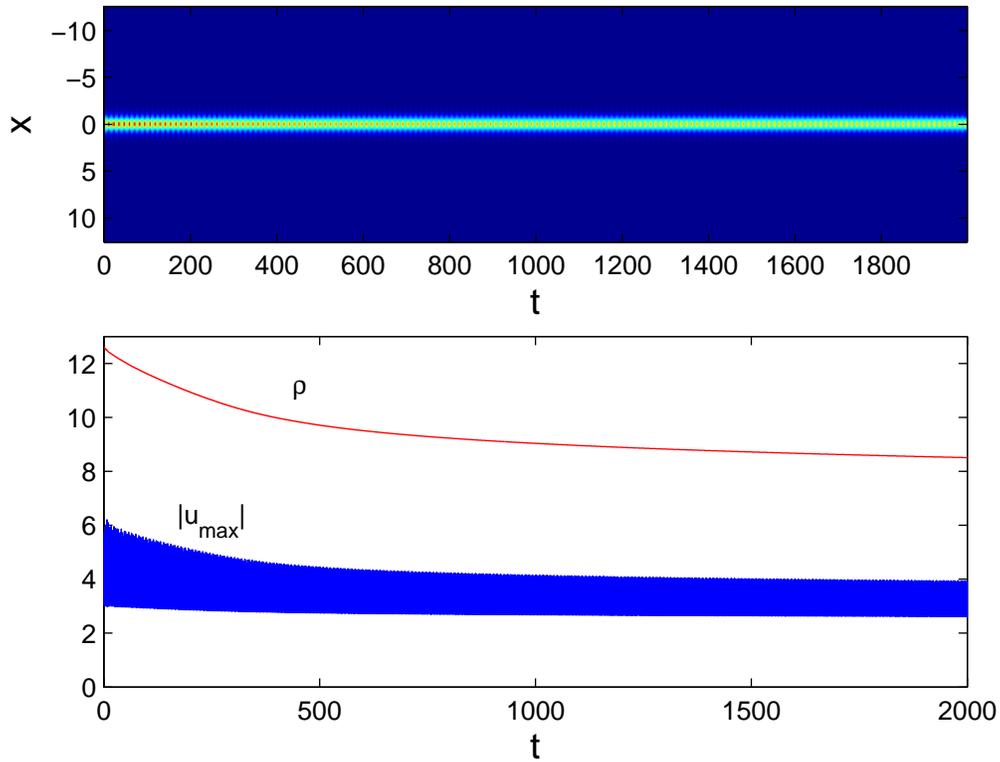}
\caption{An example of a ``semi-stable" soliton, for
$\protect\varepsilon =7.5$, $\protect\lambda _{1}=0.7$,
$\protect\omega =4.5$, and $\protect\lambda _{0}=0$, which
corresponds to point (c) in Fig. \protect\ref{fig7} (for
$\protect\lambda _{1}=0.7$). After 1400 oscillation periods, the
soliton definitely survives.} \label{fig8}
\end{figure}

\begin{figure}[tbp]
\includegraphics[scale=1]{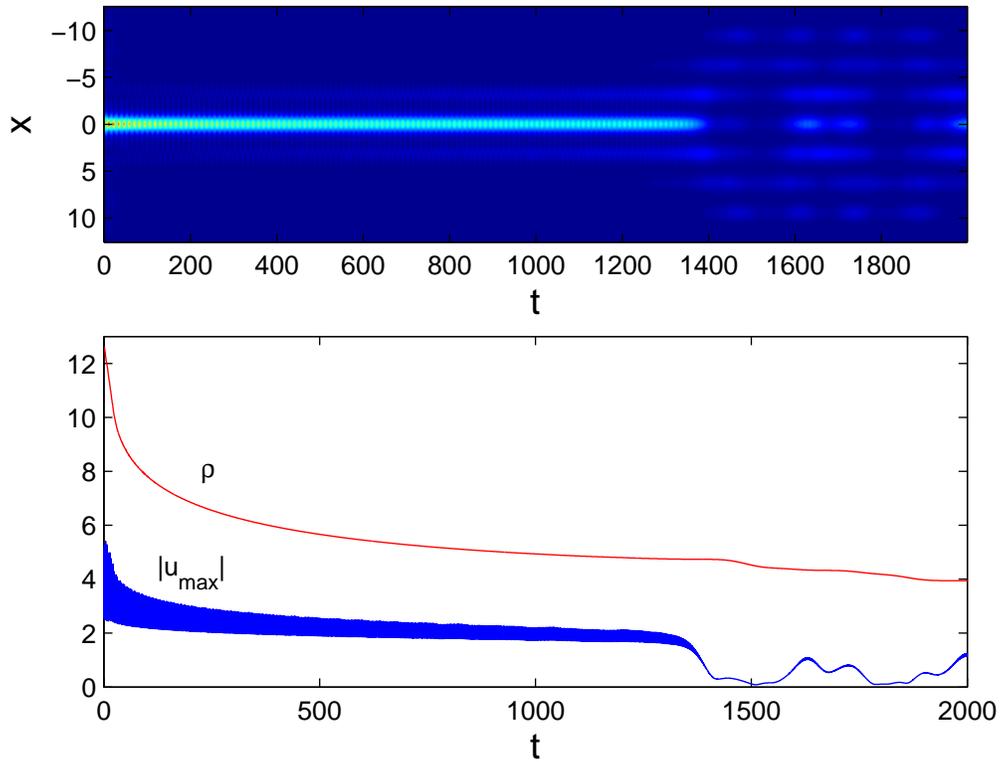}
\caption{An example of an unstable soliton, for
$\protect\varepsilon =5$, $\protect\lambda =0.7$, $\protect\omega
=4.5$, and $\protect\lambda _{0}=0$, which corresponds to point
(b) in Fig. \protect\ref{fig7} (for $\protect\lambda _{1}=0.7$).
The radiative energy loss almost ceases, but then the soliton gets
finally destroyed after about 1000 oscillation periods.}
\label{fig9}
\end{figure}

\begin{figure}[tbp]
\includegraphics[scale=1]{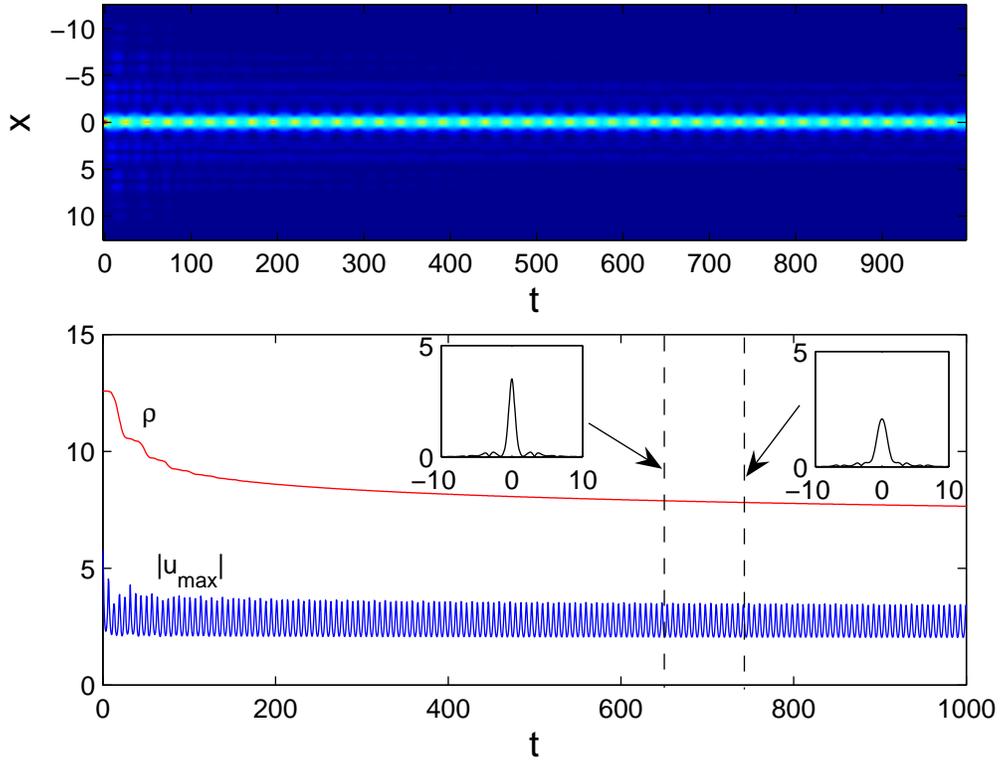}
\caption{A stable alternate soliton for $\protect\epsilon =7.5$,
$\protect\lambda _{0}=-0.9$, $\protect\lambda _{1}=1.6$, and
$\protect\omega =1$. The soliton survives while the nonlinear
coefficient periodically traverses both the $\protect\lambda =0$
point and two Bloch bands between the semi-infinite and first two
finite bands. Insets show cross-sections of instantaneous
soliton's profiles with the largest (left) and smallest (right)
amplitudes.} \label{fig10}
\end{figure}

\begin{figure}[tbp]
\includegraphics[scale=1]{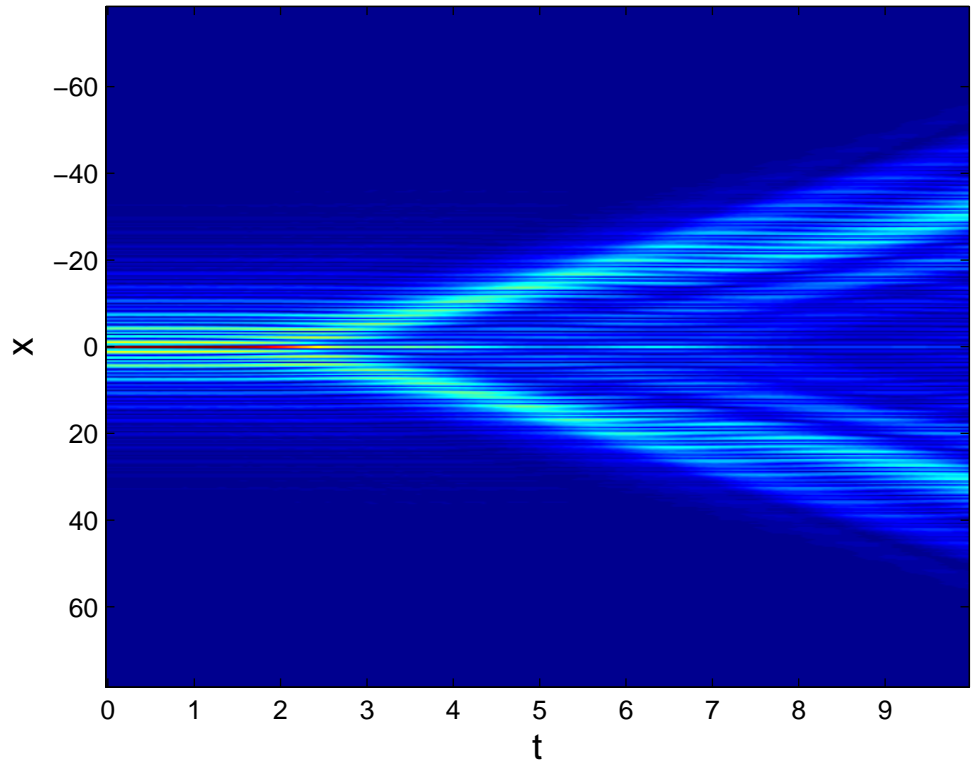}
\caption{Instability of a loosely bound soliton (with the initial
shape similar to that shown in Fig. \protect\ref{fig4}(f)), for
$\protect\varepsilon =7.5$, $\protect\lambda _{1}=-1.7$,
$\protect\omega =1$, and $\protect\lambda _{0}=0$.} \label{fig11}
\end{figure}

\begin{figure}[tbp]
\includegraphics[scale=1]{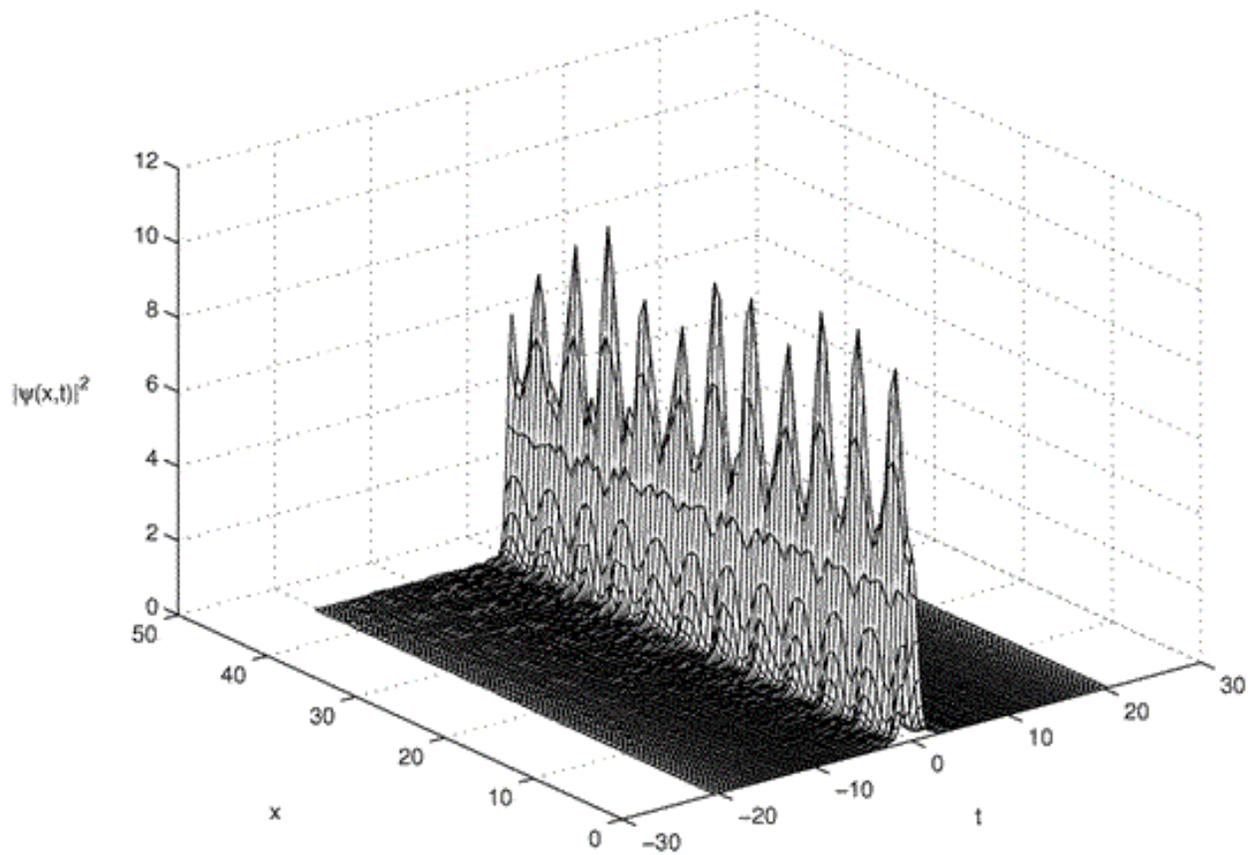}
\caption{A typical example of a stable alternate soliton found in
the one-dimensional model (\protect\ref{1D}), for
$\protect\varepsilon =4.5$, $\protect\lambda _{1}=1$,
$\protect\omega =\protect\pi /2$, and $\protect\lambda _{0}=0$.}
\label{fig12}
\end{figure}

\begin{figure}[tbp]
\includegraphics[scale=1]{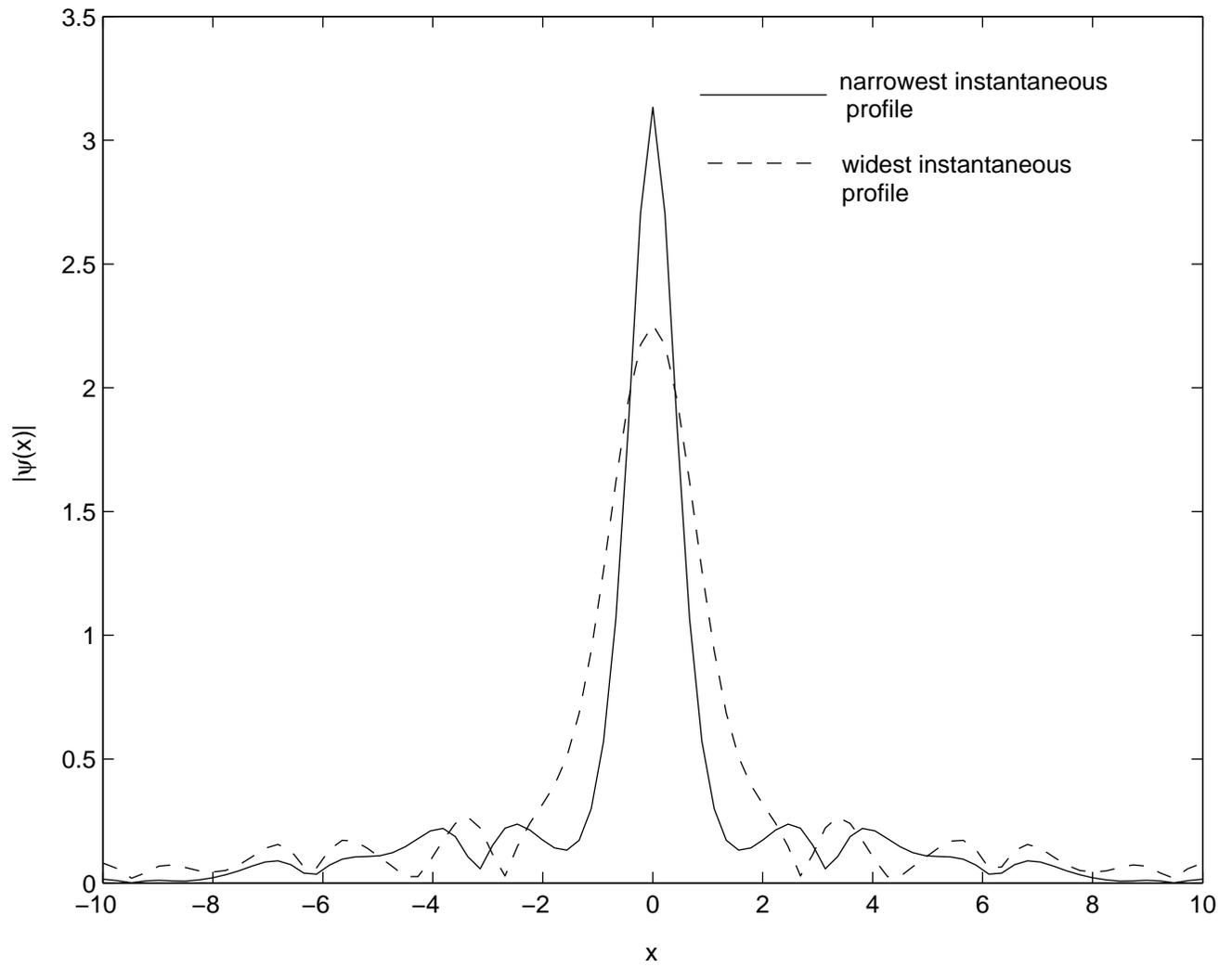}
\caption{Two profiles between which the stable alternate soliton from the
previous figure periodically oscillates.}
\label{fig13}
\end{figure}

\begin{figure}[tbp]
\includegraphics[scale=1]{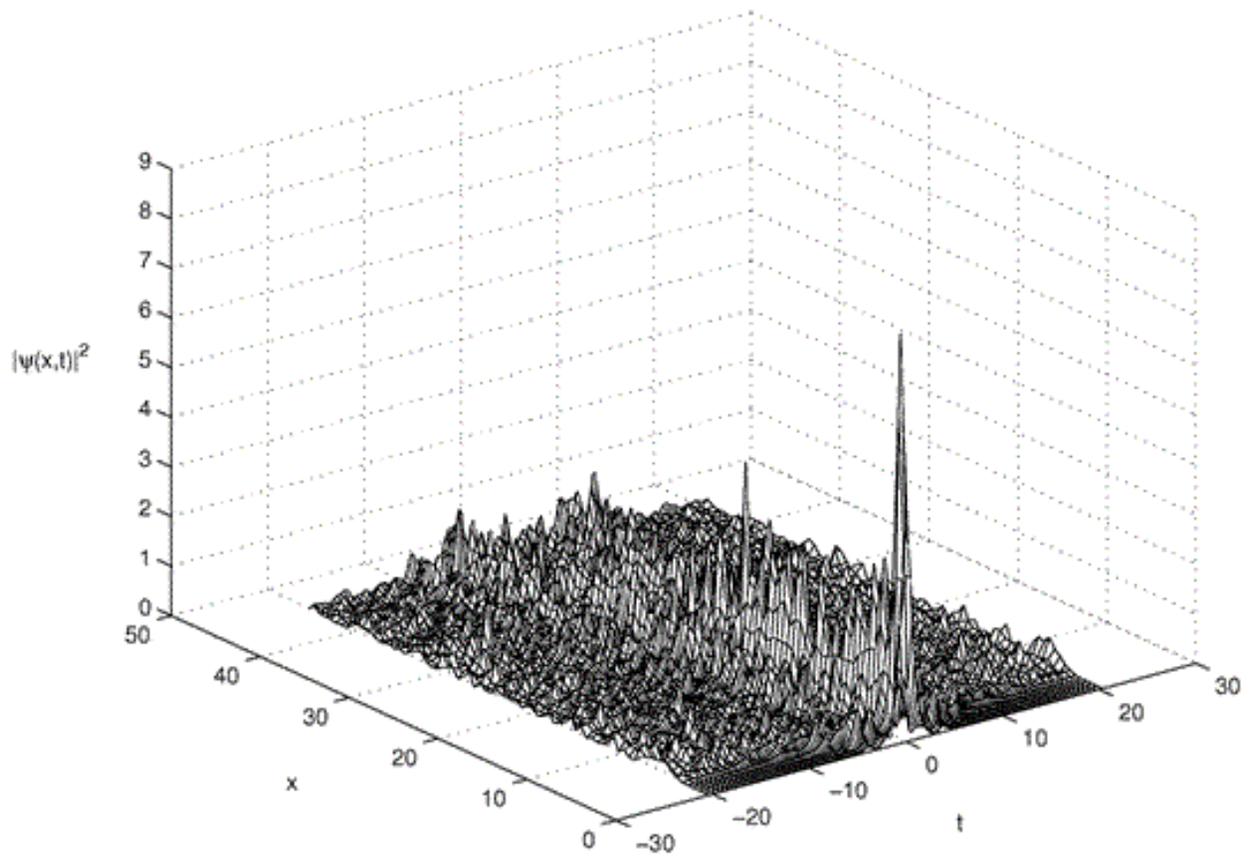}
\caption{A typical example of the instability of a one-dimensional
soliton, for $\protect\varepsilon =3.5$, $\protect\lambda _{1}=1$,
$\protect\omega =5\protect\pi /2$, and $\protect\lambda _{0}=0$.}
\label{fig14}
\end{figure}

\begin{figure}[tbp]
\includegraphics[scale=1]{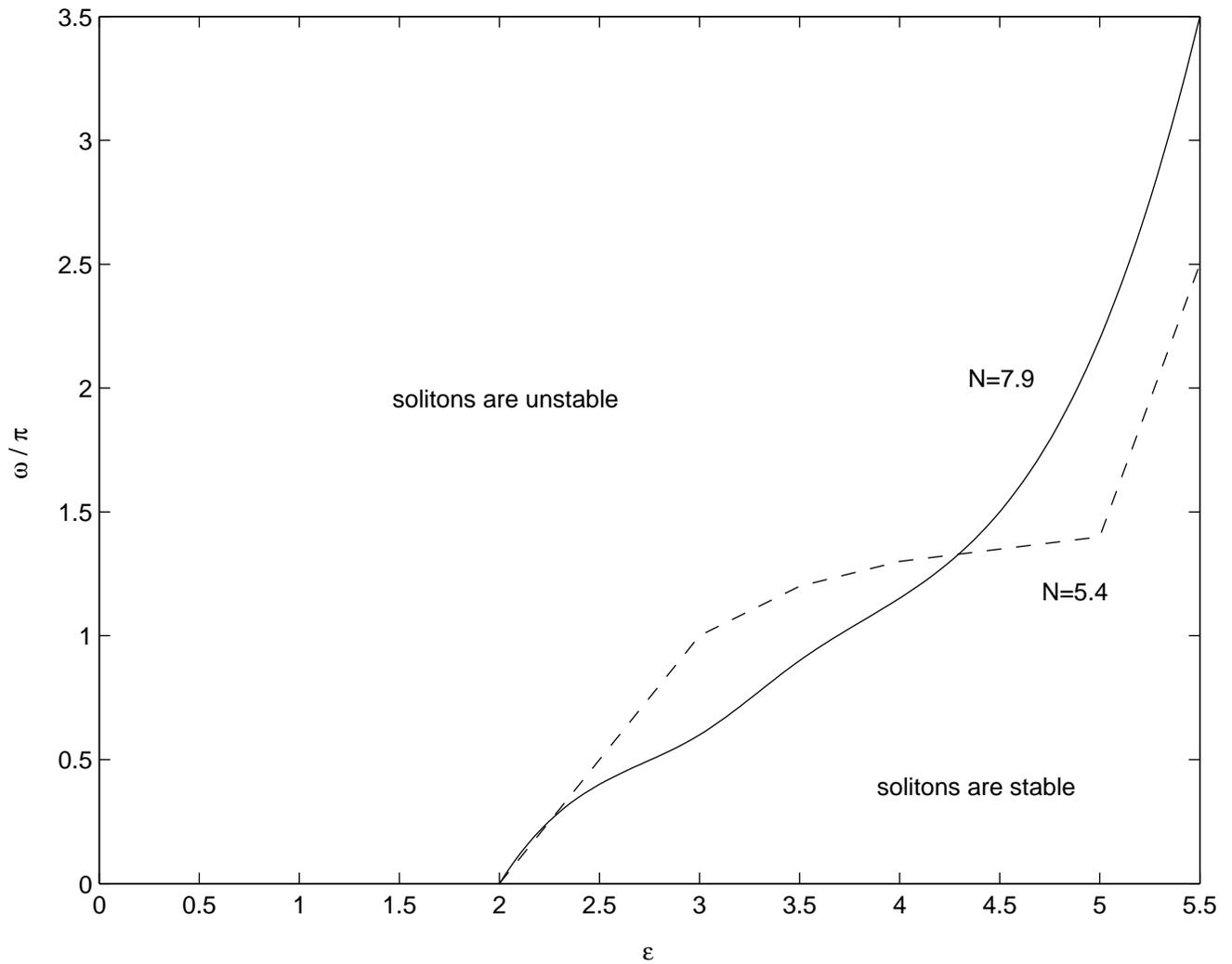}
\caption{The stability diagram for one-dimensional alternate
solitons (cf. the diagram for two-dimensional solitons in
\protect\ref{fig7} ), for $\protect\lambda _{1}=1$ and
$\protect\lambda _{0}=0$. The stability borders are shown for two
different values of the fixed norm (normalized number of atoms)
$N$, to illustrate the generality of the results.} \label{fig15}
\end{figure}

\begin{figure}[tbp]
\includegraphics[scale=1]{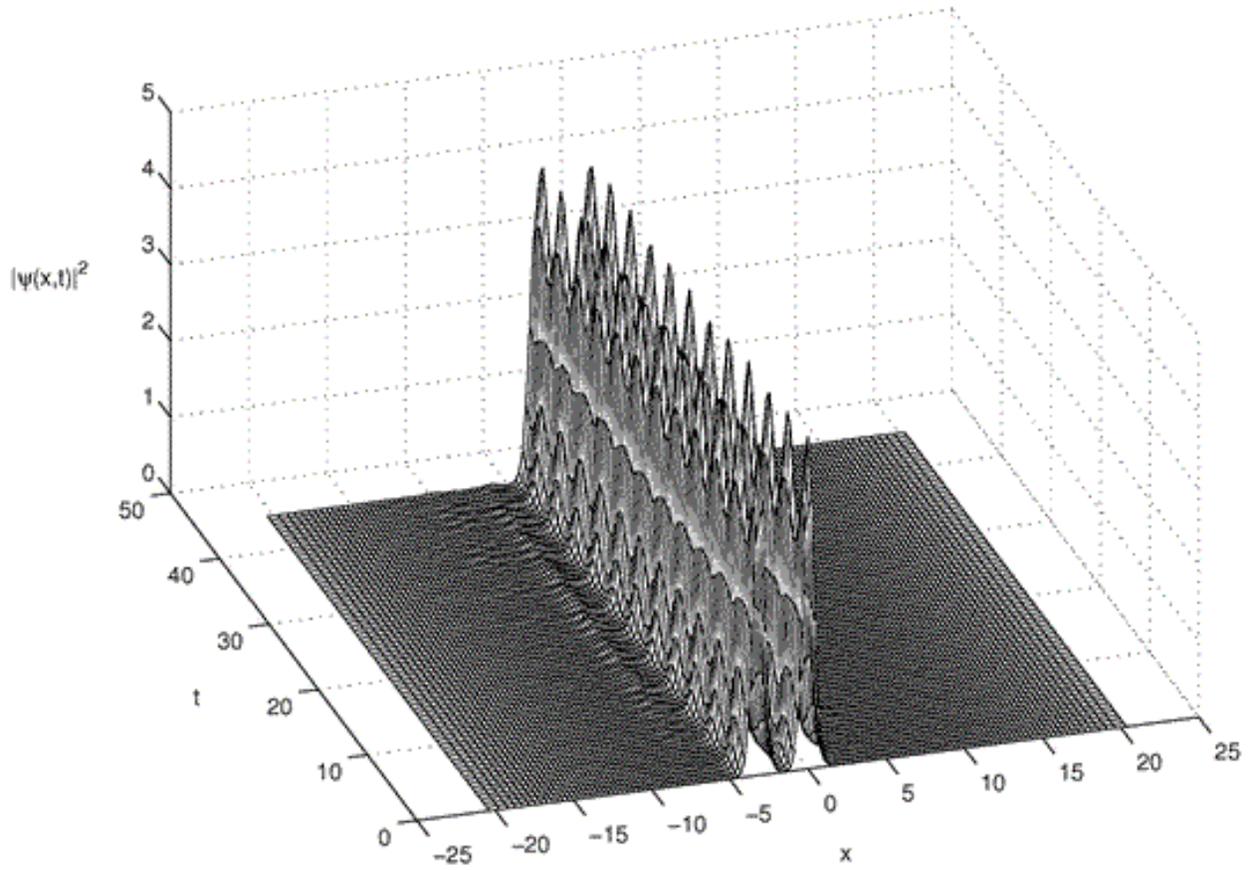}
\caption{An example of a stable odd (antisymmetric)
one-dimensional alternate soliton, for $\protect\lambda _{1}=-1$,
$\protect\varepsilon =5$, $\protect\omega =\protect\pi /2$, and
$\protect\lambda _{0}=0$.} \label{fig16}
\end{figure}
\end{subequations}

\end{document}